# Testing the energy diffusion approximation for the escape of a Brownian particle from a potential pocket

I. I. Gontchar[1], M. V. Chushnyakova[2]

[1]Physics and Chemistry Department, Omsk State Transport University, Marxa 35, Omsk, Russia 644046
[2]Physics Department, Omsk State Technical University, Mira 11, Omsk, Russia 644050

Corresponding author:
e-mail: maria.chushnyakova@gmail.com

ORCID:
0000-0002-9306-6441 (IIG)
0000-0003-0891-3149 (MVC)

**Abstract** For the first time, the energy diffusion approximation is confronted at the percent level with the exact numerical modeling of thermal decay of a metastable state. The latter is performed using the quasistationary decay rates resulting from the Langevin equations for the coordinate and conjugated momentum. For the energy (or action) diffusion approach, a Langevin-type equation for the action is constructed, validated, and solved numerically. The comparison of two approaches is performed for four potentials (two of which are anharmonic) in a wide range of two dimensionless scaling parameters: the governing parameter $G$ reflecting how high is the barrier with respect to the temperature and the damping parameter $\varphi$ expressing the friction strength. It turns out that the action diffusion approach produces the rate which is in 50% agreement with the exact one only at $\varphi < 0.02$ contrary to $\varphi < 1$ as claimed in the literature.

## 1 Introduction

Beginning from the pioneering Kramers paper [1], an escape of a Brownian particle from a potential well is the model which is useful in many branches of natural sciences from biology [2–4] to nuclear physics [5–8]. This phenomenon is considered usually in different ways depending on the friction strength. For extremely weak friction (the case of special interest in studies of the Josephson junctions [9,10] and nanowires [11]), the energy dissipation during one bounce can be neglected and the decay process is described by the approximate Kramers Formula (KF) [1]. The latter was obtained from the energy or action diffusion equation derived in [1, Eq. (14)]

$$\frac{\partial g}{\partial t} = \beta \frac{\partial}{\partial I}\left(Ig + \theta I \frac{\partial g}{\partial E}\right). \tag{1}$$

Here $q$ denotes the coordinate of the Brownian particle; $U(q)$ is the potential energy; $q_L$ ($q_R$) corresponds to left (right) turning point. The damping parameter $\beta$ of Eq. (1) is related with the friction and inertia parameters $\eta$ and $m$ as $\beta = \eta m^{-1}$.

Equation (1) is a diffusion equation in the "space" where either energy or action plays a role of the generalized coordinate; that is why it is called the energy (or action) diffusion approach. Equation (1) is approximate; it has been derived from the Fokker-Planck equation for the probability density in the phase space $P(q,p,t)$:

$$\frac{\partial P}{\partial t} = -\frac{\partial}{\partial q}\left\{\frac{p}{m}P\right\} + \frac{\partial}{\partial p}\left\{\left(\frac{\eta}{m}p + \frac{dU}{dq} + \eta\theta\frac{\partial}{\partial p}\right)P\right\}. \quad (3)$$

It is natural referring to the approach based on Eq. (3) as to the phase space approach.

The applicability of Eq. (3) is not restricted with respect to the friction strength whereas Eq. (1) is believed to be valid provided

$$\varphi = \frac{\eta\tau_c}{2\pi m} < 1. \quad (4)$$

We refer to $\varphi$ as to the damping parameter henceforth. The period of the particle oscillations near the bottom of the potential well is denoted as $\tau_c$; the corresponding frequency is $\omega_c$.

In the Kramers problem, one is interested in the quasistationary decay rate which results e.g. from Eq. (1) or from Eq. (3). From Eq. (1) Kramers obtained an approximate analytical formula [1] for the quasistationary rate

$$R_K = \omega_c \varphi G \frac{I_b}{U_b \tau_c} \exp(-G). \quad (5)$$

Here we introduce the dimensionless governing parameter

$$G = \frac{U_b}{\theta}. \quad (6)$$

$U_b$ is the height of the barrier restricting the metastable state; $I_b = I(U_b)$.

In the present work we focus on three questions: (i) at what values of $\varphi$ and how much the approximate quasistationary decay rate resulting from the action diffusion approach, $R_{DI}$, deviates from the exact rate which results from the phase space approach, $R_{DPS}$, (ii) what is the amount of agreement between $R_{DI}$ and $R_K$, and (iii) to what extent the answers depend upon the shape of the potential.

**2 The Model**

Basing on our experience [8,12,13], we model the decay process using the Langevin equations. For the exact phase space approach, these equations are equivalent to Eq. (3); they are presented in many papers (see e.g. [6,7,12,13]). In the energy diffusion approach, there is a difficulty because both $E$ and $I(E)$ enter Eq. (1) explicitly. After some transformations Eq. (1) takes the following form

$$\frac{\partial g}{\partial t} = \beta\frac{\partial}{\partial I}\left(Ig - I'\theta g + \theta\frac{\partial(gII')}{\partial I}\right). \quad (7)$$

Here $I' = dI/dE$ (note that this derivative should be expressed as a function of action). The corresponding Langevin equation for the stochastic variable $I$ reads

$$dI = -\beta(I - \theta I')dt + (\beta\theta I'I)^{1/2}\,dW, \quad (8)$$

where $dW$ is the Wiener process with the variance $2dt$. We perform numerical modeling of Eq. (8) using the Euler-Maruyama method [14].

In the modeling designed to obtain the decay rate, the initial value of the action is $I_0 = 0.01\theta\tau_c$. The modeling proceeds until either the predefined time interval $t_D$ is expired or the action becomes larger than the value $I_b$ which plays a role of the absorptive border.

In the literature the term "Langevin equations" is used usually for the stochastic differential equations describing the motion of a Brownian particle in the phase space [6,7,9,15]. We are writing and solving numerically a Langevin-type equation (8), i.e. using the action $I$ as a stochastic variable. In order to distinguish between these two types of the Langevin equations let us refer to those as to the Phase Space Langevin Equations (PSLEs) and to the Action Langevin Equation (ALE).

The modeling is performed for four potentials. The first one is the harmonic oscillator or edge potential

$$U_H(q) = \begin{cases} C_H(q - q_c)^2/2 & \text{at } q < q_b; \\ C_H(q - 2q_b + q_c)^2/2 & \text{at } q > q_b. \end{cases} \quad (9)$$

This type of potential is considered in the original Kramers paper (see [1], Fig. 2). Note that the shape of the potential beyond the barrier does not matter in the action diffusion regime due to the absence of backscattering [13].

The second potential is represented by two parabolas of the same stiffness $C_P = 4U_b/(q_b - q_c)^2$ smoothly jointed at $q_m = (q_b + q_c)/2$ ("parabolic potential"):

$$U_P(q) = \begin{cases} C_P(q - q_c)^2/2 & \text{at } q < q_m; \\ U_b - C_P(q - q_b)^2/2 & \text{at } q > q_m. \end{cases} \quad (10)$$

Potential of this shape is used in many works (see, e.g. [1,6,13,16,17]).

The third potential is borrowed from Ref. [9] ("Büttiker potential"). It reads

$$U_B(q) = U_b(1 - \cos\psi) - U_1\psi + U_2. \quad (11)$$

Here $\psi = \pi(q - q_1)$, the constants $U_1$, $U_2$, and $q_1$ were chosen to get the potential shape closer to $U_P(q)$.

The last potential is represented by the cubic parabola ("cubic potential")

$$U_C(q) = \sum_{i=0}^{3} U_{Ci} q^i. \quad (12)$$

The coefficients $U_{Ci}$ are obtained from the following conditions

$$U_C(q_b) = U_b, \quad U_C(q_c) = 0, \quad \left.\frac{dU_C}{dq}\right|_{q=q_c} = \left.\frac{dU_C}{dq}\right|_{q=q_b} = 0. \quad (13)$$

This potential was used in Refs. [17,18].

In Fig. 1 we present these four potentials (panel a), as well as the corresponding actions (panel b), and the derivatives $dI/dE$ (panel c). For convenience, in all cases the quantities are normalized: $U(q)$ is divided by $U_b$; $I(E)$ is divided by $\tau_c U_b$ whereas its argument is divided by $U_b$; the derivative $dI/dE$ is divided by $\tau_c$ whereas its argument is $I/(U_b\tau_c)$. Although even in this reduced representation the potentials in Fig. 1(a) look rather different, the dependencies of the action upon the energy are very similar to a straight line known for the harmonic oscillator ($U_H$): $I_H = \tau_c E$ (Fig. 1(b)). Unexpectedly, the reduced actions (Fig. 1(b)) and derivatives (Fig. 1(c)) for $U_C$ and $U_B$ are barely distinguishable. We are not aware of any crossing of the scientific activity of the authors of Refs. [17,18] on one hand and of Ref. [9] on the other hand. This occasional coincidence implies that the corresponding rates should be very close to each other, too.

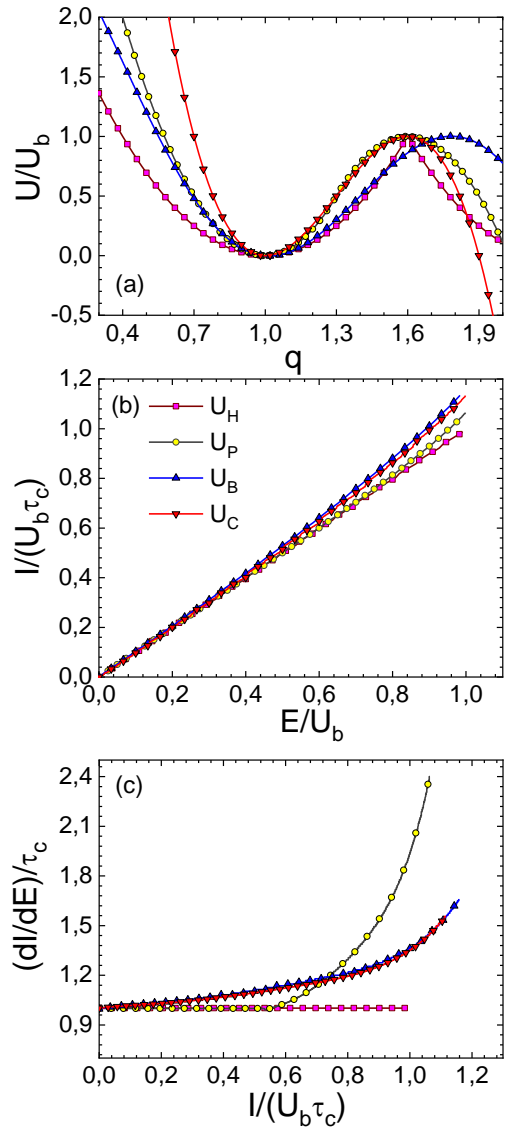

**Fig. 1** a) The potentials $U_H$, $U_P$, $U_B$, and $U_C$, divided by the barrier height $U_b$ versus the coordinate; b) the corresponding actions divided by $U_b \tau_c$ versus the energy divided by $U_b$; and c) the derivative $dI/dE$ divided by $\tau_c$ versus the action divided by $U_b \tau_c$.

Since Eq. (8) seems to absent in the literature, we first validate it in the equilibrium situation. The normalized stationary solution of Eq. (8) (or Eq. (7)) with zero flux reads

$$g_{eq} = \theta^{-1} \exp(-EG/U_b) \qquad (14)$$

resulting in $\langle E/U_b \rangle = G^{-1}$. We performed numerical modeling for $U_H$ at several values of the governing parameter using Eq. (8). Resulting distributions of the events with respect to $E/U_b$ are shown in Fig. 2 as histograms. The analytical distributions $g_{eq}$ adjusted to the numerical ones at the maximum are displayed by lines. The last column in the histograms combines all the events with $E > U_b$, i.e. the decay events. When the governing parameter decreases, the number of the decay events grows and accordingly the area of gaps under the lines corresponding to the equilibrium distribution does so.

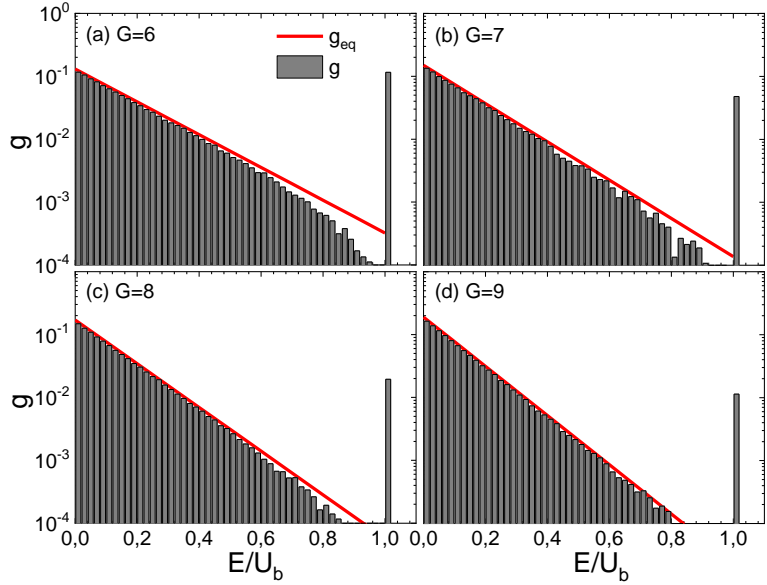

**Fig. 2** Distributions of the events with respect to $E/U_b$ (histograms) and analytical distributions adjusted to the numerical ones at the maximum (lines) for four values of $G$ displayed in the figure. The modeling has been performed using the ALE (Eq. (8)).

We see a good agreement between the two distributions at each value of $G$ for low values of $E/U_b$. This is a proof that our ALE is correct. In fact, for $G = 9$ the numerical distribution agrees with the equilibrium up to $E/U_b = 0.8$: the decay process influences the distribution only in the immediate vicinity of the barrier. As the governing parameter decreases, the domain of the argument being influenced by the decay process becomes wider.

Quantitative validation of Eq. (8) is presented in Table 1. Here we compare the value of the inverse governing parameter predefined for the modeling with $\langle E/U_b \rangle$ resulting from the modeling. This comparison confirms the validity of Eq. (8).

**Table 1** The comparison of the inverse governing parameter predefined for the modeling with $\langle E/U_b \rangle$ resulting from the modeling of the ALE for the harmonic potential; the ratio is determined within the absolute error $\Delta\langle E/U_b \rangle$.

| $G$ | $G^{-1}$ | $\langle E/U_b \rangle$ | $\Delta\langle E/U_b \rangle$ |
|---|---|---|---|
| 9  | 0.1109 | 0.1109 | 0.0026 |
| 10 | 0.1000 | 0.1003 | 0.0022 |
| 11 | 0.0909 | 0.0911 | 0.0020 |
| 12 | 0.0834 | 0.0835 | 0.0019 |
| 13 | 0.0769 | 0.0770 | 0.0017 |
| 14 | 0.0714 | 0.0715 | 0.0016 |

**3 Results**

The modeling results in $N_{tot}$ trajectories which allows finding the time-dependent decay rate

$$R_a(t) = \frac{1}{N_{tot} - N_{at}} \frac{\Delta N_{at}}{\Delta t}. \tag{15}$$

Here $N_{at}$ is the number of trajectories arriving at the absorptive border by the time moment $t$; $\Delta N_{at}$ is the number of trajectories attaining this border during the time interval $\Delta t$. Examples of the time-dependent rates obtained with the PSLEs can be found in many papers (see, e.g. [8,13,19]). As far as we know the time-dependent rates resulting from the ALE, as well as this equation itself, were not published before. Therefore, we present in Fig. 3 typical $R_a(t)$-dependences obtained both using the ALE and PSLEs for the four potentials. Note, that the damping coefficient $\beta$ can be eliminated from the ALE by renormalizing the time. Thus, it is sufficient to solve Eq. (8) only for one value of $\beta$ and then to scale the rate as follows

$$R_{I2} = \frac{\beta_{I2} R_{I1}}{\beta_{I1}}. \tag{16}$$

Here $\beta_{I1}$ ($\beta_{I2}$) are the values of the damping coefficient corresponding to the rates $R_{I1}$ ($R_{I2}$) resulting from the ALE Eq. (8). Due to the scaling property expressed by Eq. (16), we present the rates in Fig. 3 as $R/\beta$ although the PSLEs, in general, do not possess this property. The scaled rates resulting from the PSLEs and ALE, for these particular $G = 3.5$ and $\varphi \approx 10^{-3}$, look very similar for each potential. Moreover, even for different potentials the rates are hardly distinguishable: in all four panels of Fig. 3, they reach more or less the same quasistationary value $R_D/\beta \approx 90 \cdot 10^{-3}$ after a relaxation stage with the duration $\tau_r \beta \approx 2$. Although the dynamical rates in Fig. 3 significantly fluctuate, we manage to find the value of $R_D$ with the relative error $\varepsilon_D < 2\%$ (typically 1%).

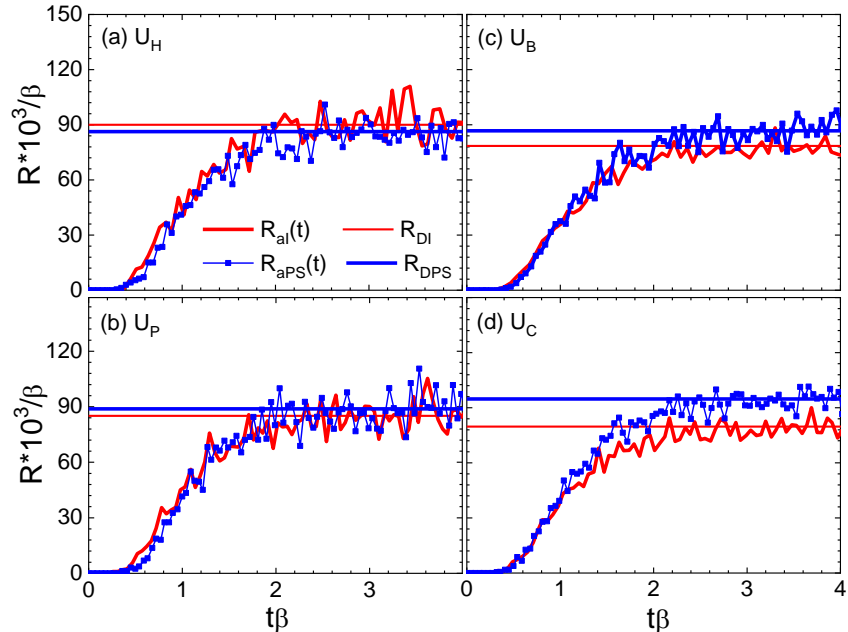

**Fig. 3** The scaled time-dependent rates (wriggling lines) and the corresponding quasistationary rates (horizontal lines) for four potentials under consideration. $G = 3.5$, $\varphi \approx 10^{-3}$.

Let us now compare systematically the quasistationary numerical rates resulting from the ALE, $R_{DI}$, with the approximate Kramers rates, $R_K$ (Eq. (5)). Note that the value of $I_b$ entering this equation is calculated numerically. In Fig. 4(a) we present the rates $R_{DI}$ divided by the damping coefficient versus

$G$. In the considered domain of the governing parameter the rates cover nearly 4 orders of magnitude and behave with $G$ qualitatively as predicted by the Kramers formula, Eq. (5). The quantitative comparison is made in Fig. 4(b),(c) where the ratio $r_{IK} = R_{DI}/R_K$ is presented. At $G < 1.5$ for all four potentials the Kramers formula significantly underestimates the quasistationary dynamical rate resulting from the ALE, i.e. within the same action diffusion approach. This is expected because the low barrier does not provide enough time for the probability density to relax in the potential pocket (although the value $r_{IK} \approx 2$ at $G = 1$ seems to be unexpectedly high basing on our experience in the domain of the overdamped motion [8]). At $G > 3$ for all four potentials the Kramers formula significantly overestimates the dynamical rate. However, in this domain of the governing parameter, for the harmonic potentials ($U_P(q)$ and $U_H(q)$) the Kramers formula is in better agreement with $R_{DI}$ deviating by 20 to 10%. The anharmonic potentials result in $r_{IK} \approx 0.6$ with no a tendency to increase.

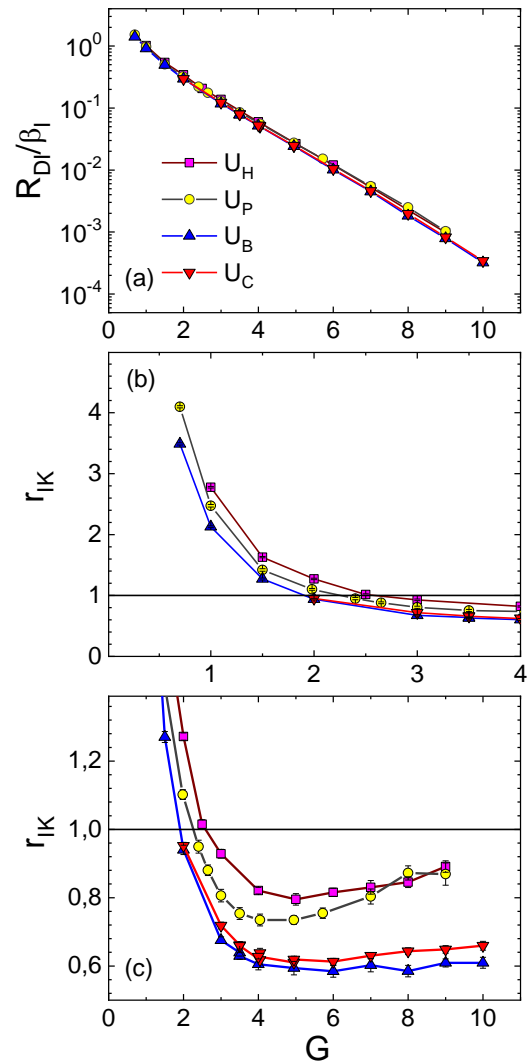

**Fig. 4** Versus the governing parameter the following quantities are shown for four considered potentials: a) The scaled quasistationary dynamical rates $R_{DI}/\beta_I$ ; b) and c) the ratio $r_{IK} = R_{DI}/R_K$ in different scales.

The dynamical rates resulting from the approximate action diffusion approach, $R_{DI}$, are compared with the exact rates $R_{DPS}$ in Figs. 5 and 6. For this aim we present the ratio

$$r_{IPS} = \frac{R_{DI}\beta_{PS}}{R_{DPS}\beta_I}. \tag{17}$$

Here $\beta_{PS}$ ($\beta_I$) is the value of the damping coefficient with which the PSLEs (ALE) are employed. First, in Fig. 5 we see that $r_{IPS}(G)$ dependencies look very similar for the $U_H(q)$ (Fig. 5(a)) and $U_P(q)$ (Fig. 5(b)) potentials. This is unexpected remembering the very different behavior of $dI/dE$ for these two potentials in Fig. 1(c)). The similarity suggests that this is the vicinity of the potential minimum that plays a decisive role for the rate, not the shape of the barrier.

In Fig. 6 the same ratio $r_{IPS}$ is presented as a function of $\varphi$ for $G = 3.5$ and 5.0. First, the $\varphi$-dependencies look similar in panels (a) and (b): this is in agreement with Fig. 5. Second, the four dependencies group by two: the ratios obtained using $U_H$ and $U_P$ form one group whereas the ratios resulting from $U_B$ and $U_C$ are comprised in another group. This again correlates with the behavior of the derivative $dI/dE$ in Fig. 1(c), although this derivative influences only $R_{DI}$, not $R_{DPS}$. As expected, the ratio $r_{IPS}$ becomes close to the unity at small values of $\varphi$, though the large value $r_{IPS} \approx 2$ at small $\varphi \approx 0.05$ is somewhat surprising.

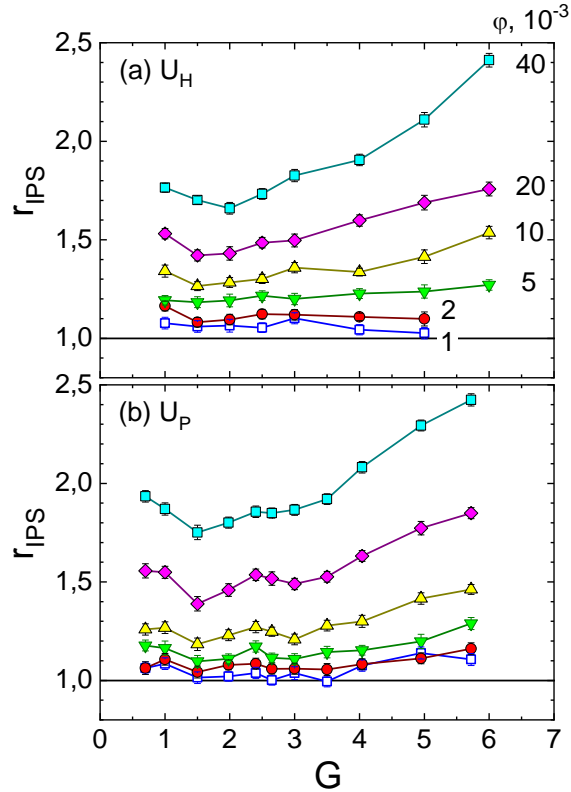

**Fig. 5** The ratio $r_{IPS}$ as a function of the governing parameter at 6 values of the damping parameter for the harmonic (a) and parabolic (b) potentials.

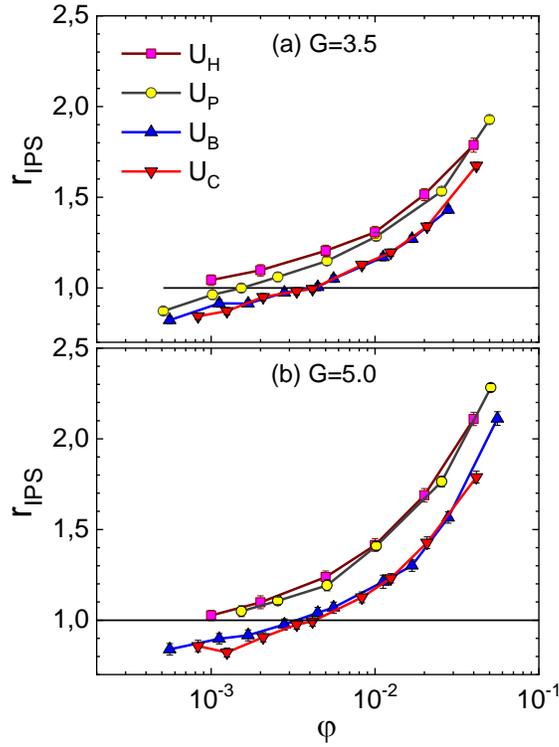

**Fig. 6** The ratio $r_{IPS}$ as a function of the damping parameter at 2 values of the governing parameter for four considered potentials.

## 4 Summary and conclusions

The energy (action) diffusion approximation for the first time has been tested at the percent level using the numerical modeling of thermal decay of a metastable state. This modeling is considered to give the exact rates within the statistical errors (not exceeding 2% in the present study). The exact modeling has been performed using the Langevin equations for the coordinate and conjugated momentum (PSLEs). For the action diffusion approach, a Langevin-type equation (ALE) has been derived for the action as a stochastic quantity. This equation has been tested by means of the comparison with the analytical equilibrium distribution and then solved numerically. The two approaches have been confronted for four substantially different potentials in a wide range of two dimensionless parameters $G$ and $\varphi$ (see Eqs. (4) and (6) for their definitions).

As the first step, we have compared the approximate Kramers rate $R_K$ (see Eq. (5)) with the numerical rate obtained using the ALE, $R_{DI}$. The latter rate is exact within the action diffusion approach. In general, the agreement of $R_K$ with $R_{DI}$ is not good: the difference is about 30% at $G > 3$ where much better agreement might be expected.

Then $R_{DI}$ has been compared with the exact numerical rate resulting from the more accurate PSLEs, $R_{DPS}$. It turns out that the rate resulting from the action diffusion approach is in 50% agreement with $R_{DPS}$ only at $\varphi < 0.02$ whereas in the literature usually $\varphi < 1$ is stated as the condition of the applicability of the action diffusion approach.

Due to the scaling property Eq. (16), the ALE requires significantly less computer resources than the PSLEs. Using results of the present study might be useful to others saving the computer time and avoiding inaccuracies related with the implementation of the faster action diffusion approach.